\documentclass[]{interact}

\usepackage{epstopdf}
\usepackage[caption=false]{subfig}
\usepackage[breakable]{tcolorbox}
\usepackage{xcolor}
\usepackage{algorithm}
\usepackage{algpseudocode}
\usepackage{comment}
\usepackage{listings}
\usepackage[numbers,sort&compress]{natbib}
\usepackage{tikz}
\usetikzlibrary{arrows.meta, positioning, fit, backgrounds, shapes.geometric}
\bibpunct[, ]{[}{]}{,}{n}{,}{,}

\theoremstyle{plain}

\theoremstyle{definition}

\theoremstyle{remark}

\newtcolorbox{llmprompt}{
colback=gray!10,
colframe=gray!50!black,
boxrule=1pt,
arc=4pt,
title=Instruction Prompt,
fonttitle=\bfseries,
breakable
}

\begin{document}

\articletype{ARTICLE}

\title{Metadata Augmentation using NLP, Machine Learning and AI-chatbots: A comparison}

\author{
\name{Alfredo Gonz\'alez-Espinoza\textsuperscript{a*}\thanks{*Email: agonzal3@andrew.cmu.edu}, Dom Jebbia\textsuperscript{a} and Haoyong Lan\textsuperscript{a}}
\affil{\textsuperscript{a} University Libraries, Carnegie Mellon University, Pittsburgh, Pennsylvania, USA}
}

\maketitle

\begin{abstract}
Recent advances in machine learning and artificial intelligence have provided more alternatives for the implementation of repetitive or monotonous tasks. However, the development of AI tools has not been straightforward, and use case exploration and workflow integration are still ongoing challenges. In this work, we present a detailed qualitative analysis of the performance and user experience of popular commercial AI chatbots when used for document classification with limited data. We report the results for a real-world example of metadata augmentation in academic libraries environment. We compare the results of AI chatbots with other machine learning and natural language processing methods such as XGBoost and BERT-based fine tuning, and share insights from our experience. We found that AI chatbots perform similarly among them while outperforming the machine learning methods we tested, showing their advantage when the method relies on local data for training. We also found that while working with AI chatbots is easier than with code, getting useful results from them still represents a challenge for the user. Furthermore, we encountered alarming conceptual errors in the output of some chatbots, such as not being able to count the number of lines of our inputs and explaining the mistake as ``human error''. Although this is not complete evidence that AI chatbots can be effectively used for metadata classification, we believe that the information provided in this work can be useful to librarians and data curators in developing pathways for the integration and use of AI tools for data curation or metadata augmentation tasks.  
\end{abstract}


\section{Introduction}

Metadata augmentation is a task that librarians and repository managers use to provide findability and accessibility of digital objects, and ensure consistency of metadata across platforms. This process can often be systematically addressed when new or missing fields are available elsewhere. However, in cases where metadata fields are missing or unknown, alternative approaches that leverage machine learning techniques may be employed.

Methods based on natural language processing (NLP) and machine learning are regularly used for metadata curation and augmentation \cite{jiang_deep_2024,safder_deep_2020,duarte_review_2023}. These techniques have shown promise in extracting relevant information from unstructured text to enhance metadata quality \cite{trivedi_automatic_2018}.

In recent years, the emergence of large language models (LLMs) has introduced powerful new tools for complex language tasks. LLMs are neural networks trained on massive text corpora, enabling them to generate human-like text, answer questions, and assist with classification and extraction \cite{bhayana_chatbots_2024}. The development of LLMs has opened up new possibilities for exploring their performance and potential applications, such as in the form of AI-powered chatbots, for document organization or classification and metadata augmentation\cite{bayer_data_2023, bayer_survey_2022}.

Chatbots built on LLMs have demonstrated extensive capabilities across various domains. In medicine, they have shown some proficiency in answering questions, summarizing articles, and even suggesting that they match expert human performance on professional benchmarks \cite{singhal_large_2023}. In the library and information science field, early studies have explored using chatbots as auxiliary tools for cataloging tasks like generating MARC records \cite{brzustowicz_chatgpt_2023} and assigning subject headings \cite{zhang_utilising_2023}, with promising results.

This study aims to evaluate and compare the performance and user experience of traditional machine learning and NLP methods with cutting-edge AI tools, specifically LLMs and chatbots, for metadata augmentation tasks. By documenting their strengths, limitations, and potential applications, we seek to provide insights and guidance for librarians and repository managers considering the adoption of these emerging technologies to integrate them in their workflows\cite{schopfel_new_2025}.

\section{Case Study: Graduate Theses Metadata}
KiltHub, Carnegie Mellon University's institutional repository, has hosted various research outputs since 2018, including publications, datasets, presentations, software, and theses. Currently, KiltHub contains approximately 3,700 theses.

\begin{figure}[H]
    \centering
    \includegraphics[width=0.75\linewidth]{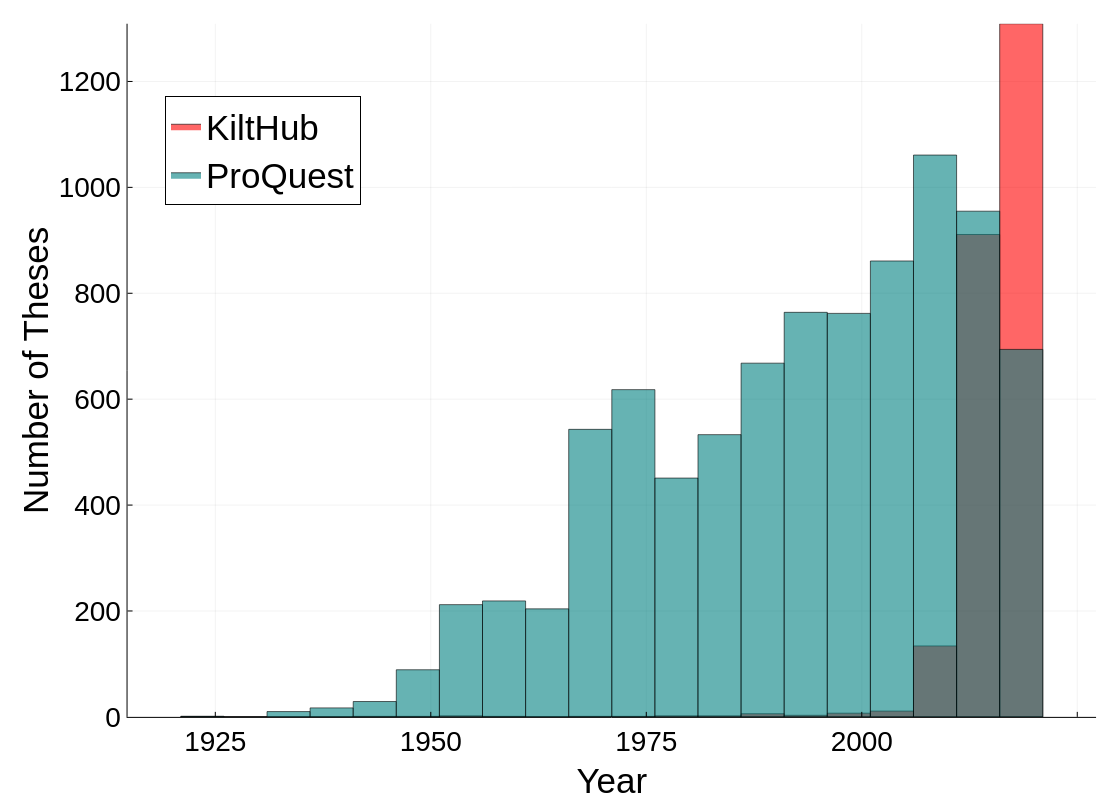}
    \caption{Number of theses and dissertations hosted on both platforms distributed by date of publication.}
    \label{fig:tnumber_years}
\end{figure}

Comprehensive tracking of student work is essential for multiple purposes, including:
\begin{itemize}
    \item Showcasing individual academic achievements
    \item Identifying research trends across departments and colleges
    \item Developing data-driven infrastructure and services
\end{itemize}
ProQuest, a major academic publishing company, maintains a significantly larger collection of CMU theses (approximately 9,300). However, around 7,800 of these theses lack departmental metadata, which also means they lack college and department affiliation information. \\
Integrating and augmenting the ProQuest metadata with college and departmental information would substantially improve the organization and findability of student work, enhancing the value of ProQuest's existing metadata.

\subsection{Adding the field ``College''}

\begin{figure}
    \centering
    \includegraphics[width=0.65\linewidth]{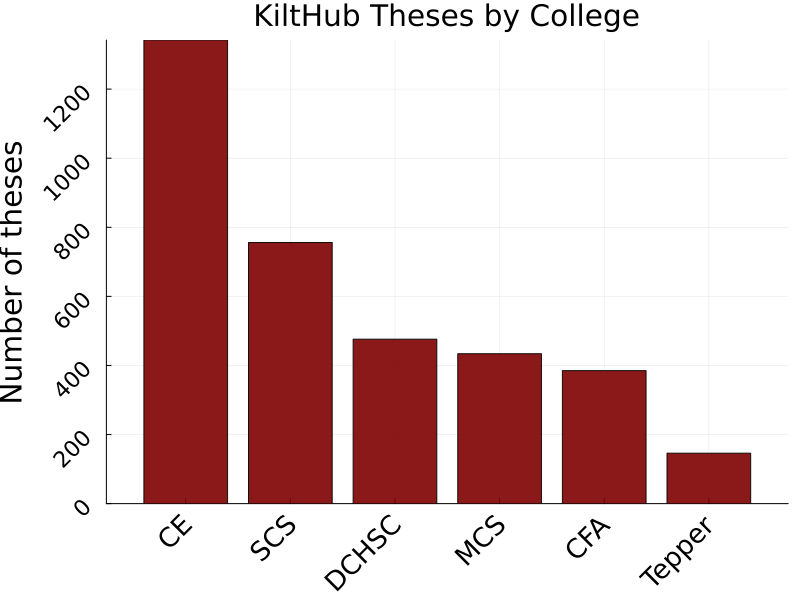}
    \caption{Number of theses by College hosted on KiltHub. Colleges are: School of Computer Sciences (SCS), Dietrich College of Humanities and Social Sciences (DCHSC), Tepper School of Business, College of Fine Arts (CFA), College of Engineering (CE) and Mellon College of Science (MCS). Heinz College of Information Systems and Public Policy is not included in the plot since it only had 13 theses.}
    \label{fig:college_distro}
\end{figure}

To create the College label for each thesis, we begin by assigning labels to entries where we have confirmed ground-truth data. This process leverages a dictionary data structure that maps colleges to their respective departments within Carnegie Mellon University's organizational structure. Each entry in our collection of theses is processed to add a new ``College'' metadata field, which is populated based on the thesis's department affiliation. For theses where department information is unavailable, the field is marked as ``missing''. The labeling process was implemented following the steps shown in Algorithm \ref{algo_college}.\\
\begin{algorithm}
\caption{Assign Colleges to Theses}
\begin{algorithmic}[1]
\Require Collection of thesis metadata records, Dictionary mapping colleges to lists of departments
\Ensure Updated thesis records with college assignments
\For{each thesis in collection}
    \State Get department from thesis metadata
    \If{department exists}
        \State Find corresponding college in mapping
        \State Assign college to thesis metadata
    \Else
        \State Mark college as ``missing''
    \EndIf
\EndFor
\State \Return Updated thesis collection
\end{algorithmic}
\label{algo_college}
\end{algorithm}
\\
After assigning new labels to our documents, we examined the class distribution across our dataset. Figure \ref{fig:college_distro} shows how documents are distributed by college, revealing a notable imbalance in sample counts across different labels. This class imbalance is an important characteristic of our dataset that needs consideration, especially when using methods that rely on training data. 

\subsection{College labeling as classification problem}
The process of assigning appropriate ``College'' metadata to university theses can be conceptualized as a subject classification task within library cataloging. In this framework, given a collection of $n$ uncataloged theses $T = \{t_1, t_2, \ldots, t_n\}$, and a controlled vocabulary of institutional colleges $C = \{c_1, c_2, \ldots, c_k\}$, we seek to develop a cataloging function $f: T \rightarrow C$ that appropriately assigns each thesis to its corresponding college.

Each thesis $t_i$ is characterized by its content representation $x_i$, where:
\begin{itemize}
    \item $x_i = [x_{i1}, x_{i2}, \ldots, x_{id}]$ constitutes the content representation for thesis $i$
    \item $d$ represents the dimensionality of the representation space (vocabulary size or embedding dimensions)
    \item Each element $x_{ij}$ indicates the significance or presence of term $j$ in thesis $i$
\end{itemize}

The content representation captures the thematic elements of the document, typically expressed through term frequency-inverse document frequency (TF-IDF) weights, keyword presence, or semantic embeddings of titles or abstracts.

The classification designation $y_i \in C$ represents the college assignment for thesis $i$, where:
\begin{itemize}
    \item $y_i = f(x_i)$ is the assigned college classification
    \item $f$ is the cataloging function we aim to develop or discover
    \item The training corpus consists of previously cataloged examples $\{(x_i, y_i)\}_{i=1}^m$ where $m$ is the number of theses with existing college assignments
\end{itemize}

Classification models learn the underlying patterns in the training data to make accurate predictions on unseen documents. By treating thesis features as inputs and college assignments as targets, various algorithms can be employed, such as logistic regression, decision trees, or neural networks.

For metadata augmentation purposes, the literal classification function $f$ is not required. Instead, the goal is to find a model that performs sufficiently well on the labeled data (e.g., from KiltHub metadata) and can be applied to theses with missing college information.

By formulating thesis college assignment as a multiclass classification problem, established machine learning techniques can be leveraged to automate and scale the metadata augmentation process. This approach has the potential to significantly reduce manual effort while maintaining accuracy in assigning documents to their respective labels.

\section{Testing traditional NLP/ML methods against AI-chatbots}

This study evaluates three distinct approaches to subject classification for theses cataloging, comparing traditional Natural Language Processing with Machine Learning techniques against emerging Large Language Models. Our analysis aims to provide insights into the reliability and resource efficiency of these methodologies for metadata assignment in academic library collections. Particular attention is given to the implementation requirements of each approach, including technical expertise and infrastructure needs within library technical services.

The comparative framework examines three methodological approaches:

1. \textbf{Natural Language Processing $+$ Machine Learning}: Combining traditional text processing techniques with machine learning classifiers.

2. \textbf{Large Language Model with Fine-tuning}: Adapting pre-trained language models through additional training on institutional thesis collections and controlled vocabulary.

3. \textbf{AI-chatbot with Prompt Engineering}: Utilizing commercial generative language models with prompt engineering

\begin{figure}[H]
    \centering
    \includegraphics[width=0.8\columnwidth]{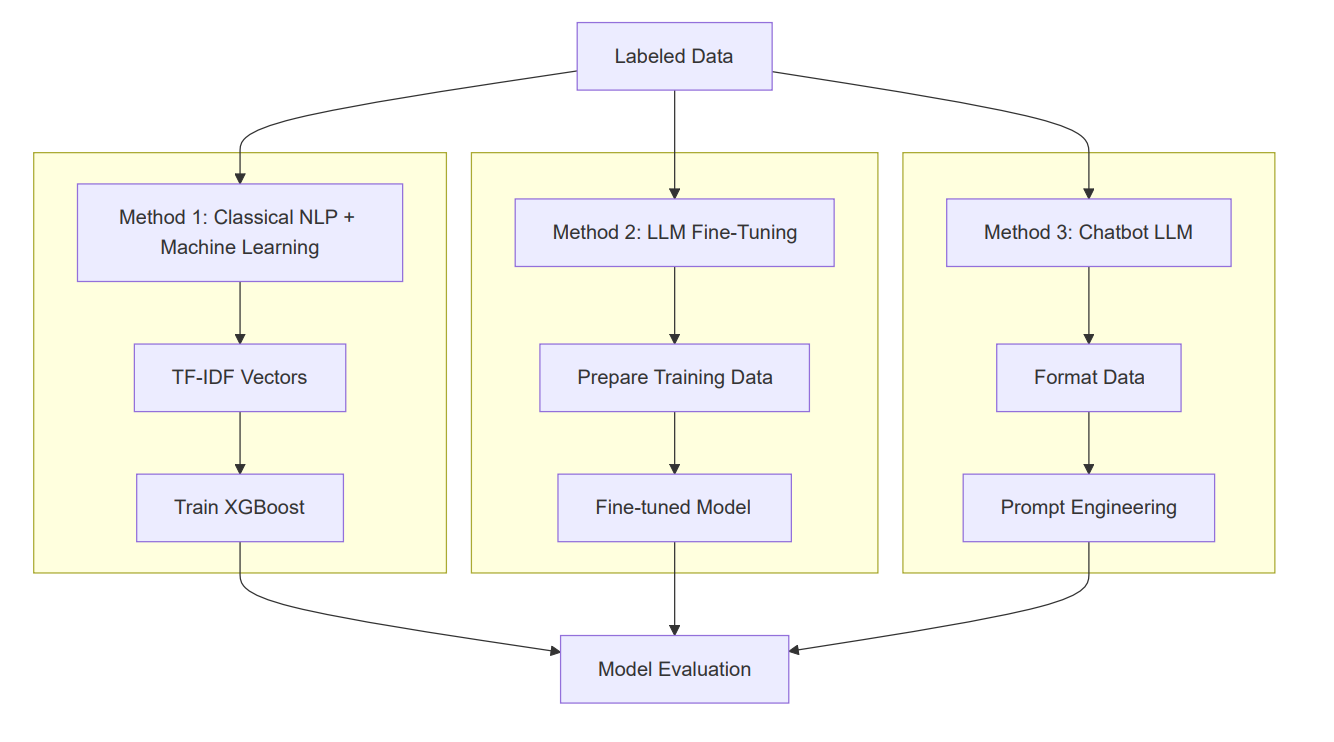}
    \caption{Workflow diagram for the three different methods implemented.}
    \label{fig:enter-label}
\end{figure}

\subsection{Method 1: Natural Language Processing + Machine Learning}
We used XGBoost \cite{chen_xgboost_2016}, a decision tree based model as our machine learning reference. XGBoost is a widely used machine learning model that has gained popularity after winning several competitions for performance with high dimensional tabular data\cite{noauthor_xgboostdemo_nodate}.

We build a Term Frequency-Inverse Document Frequency matrix to use it as features for our XGBoost model. Each document is defined as a concatenation of terms included in the title and keywords:
\begin{equation}
d_i = \text{title}_i \oplus \text{keywords}_i \text{,}
\label{eq:d_i}
\end{equation}

where the title is processed before concatenation (removing punctuation and stopwords). After creating each document, we compute the matrix:
\begin{equation}
\text{tfidf}(t,d,D) = \text{tf}(t,d) \cdot \text{idf}(t,D)\text{,}
\label{eq:tfidf}
\end{equation}

Where $\text{tf}(t,d)$ is a value that represents the normalized frequency of term $t$ in document $d$, and $\text{idf}(t,D)$ is a value that represents the number of documents with a term $t$ with respect to the total number of documents (see appendix for details).

The TF-IDF metric assigns higher weights to terms that appear frequently in specific documents while being relatively rare across the corpus. Terms appearing frequently across all documents receive lower weights, effectively filtering out common but non-distinctive language.

We performed a grid search (see appendix) to find the XGBoost parameters with best performance for the classification task.

\subsection{Method 2: Large Language Model with Fine-tuning}

We tested a BERT-based fine-tuning methodology to learn if fine-tuning a relatively small and computationally cheap large language model, such as BERT, can be used for this specific classification task. BERT is an open source model introduced by Google in 2018, and was trained using the Toronto BookCorpus and the English Wikipedia \cite{devlin_bert_2019}. BERT has been used extensively to build domain-specific LLMs and other fine-tuned applications such as sentiment analysis, document classification, and question/answer chats\cite{wang_utilizing_2024,church_emerging_2021}. Fine-tuning is a technique that involves training only a part of the Large Language Model to perform specific tasks such as text classification \cite{qasim_fine-tuned_2022}. \\

We tokenized the pre-processed text from equation \ref{eq:d_i} to prepare it for training. After tokenization, BERT uses contextual embeddings where each token $t$ is embedded based on its context $c$: $f(t|c) \rightarrow \mathbb{R}^d$, with $d=768$ for BERT-base. This embedding captures semantic relationships given by the text it was trained on, and these relationships are the ones the classifier model will identify to use as features in the fine-tune (training) process. \\

To fine tune the BERT-based model we used the Huggingface transformers API to initialize and train our classifier with the \texttt{AutoModelForSequenceClassification} pre-trained BERT model\cite{HuggingFaceTransformersTraining}. 
Each document was processed in the same way as the XGBoost classifier, using train and test sets for training and a set of unseen data for evaluation.

The complete training and testing methodology can be found in the Appendix and supplementary materials.

\subsection{Method 3: AI-chatbot}
Finally, we tested some of the popular large language models in their chatbot form. For these experiments we decided to use a zero-shot approach\cite{anglin_automatic_2024}. In the same fashion that we attempt to make use of the word representations (embeddings) created by training the LLMs, we want to implicitly use the embeddings the LLMs behind the chatbots have as features for our chatbot classifier. We generated 5 samples of 10 documents for each college (5x70 documents in total). The samples were generated via random permutation with replacement from the original dataset. \\
The prompt used for the experiments is as follows:
\begin{llmprompt}
      You are a cataloger assigning metadata terms to documents. You will be provided an array containing lists of terms separated by commas, each line representing a $list\_of\_terms$ or document to classify. Pick the college that seems most appropriate based on the terms in the list. Output a list of `labels' corresponding to the name of the college for each line.  \\
      Choose from these colleges:
\end{llmprompt}

After the paragraph we include a list of colleges with their respective departments to add  more context in the prompt.\\
We performed the experiments with Claude Sonnet 3.5 and 3.7, Microsoft Copilot, ChatGPT 4o and o4-mini, Gemini 2.0 Flash, Flash Thinking and 2.5 Pro. Some chatbots allow users to add `context' or `instructions'; for example, Claude offers a feature called `Project Knowledge'. We tested this feature and found that while it helped keep the chat organized, it did not significantly improve performance (see the full performance table in appendix).

\section{Results and discussion}
To compare results between the models we focused on two aspects: performance and user experience. We are interested in evaluating the feasibility of integrating these commercial tools into workflows with repetitive tasks.

\subsection{Performance}
\begin{figure}[H]
    \centering
    \includegraphics[width=0.95\columnwidth]{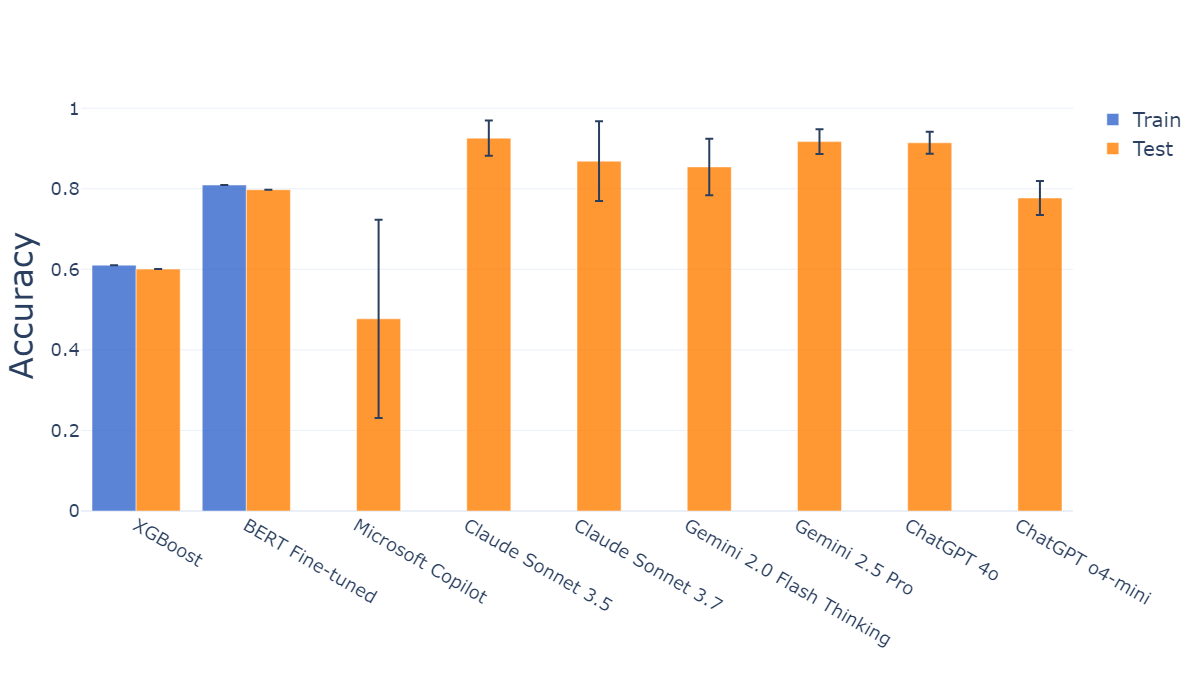}    
    \caption{Accuracy results for the models tested. Only the XGBoost and BERT fine-tuned methods were tested with train/test data given the fact that chatbots don't require training data in zero shot scenarios. Error bars correspond to the standard variation of the accuracy for the group of documents (5x70)}
    \label{fig:model_accuracy}
\end{figure}

The results for the methods and models tested are shown in Fig \ref{fig:model_accuracy}. XGBoost demonstrated relatively poor performance compared to the BERT fine-tuned method, an expected outcome given the limited vocabulary available from keywords and titles versus the extensive language understanding capabilities offered by fine-tuning a large language model. While most chatbot models performed with comparable accuracy, Microsoft Copilot exhibited the lowest accuracy and greatest inconsistency in its results, suggesting potential lack of Copilot's development and maintenance compared to other commercial LLMs. We excluded results for Gemini 2.0 Flash from our analysis because these models consistently generated an incorrect number of labels, preventing us from collecting the necessary 5×70 results required for evaluation.

The XGBoost and BERT fine-tuned models performed the worst in cases with fewer training samples: specifically Heinz and Tepper, where the evaluation data consisted of only 3 samples for Heinz and 20 for Tepper. This result confirms the limitation of models that depend heavily on training data, a constraint that appears minimal in the most advanced LLM chatbots. Although chatbots generally perform better with labels such as Science or Engineering, some experiments showed lower accuracy for the Heinz and Tepper labels. However, they still outperform the sample size-dependent methods (see Table \ref{tab:model-comparison}).

\begin{table}[htbp]
\centering
\small
\begin{tabular}{|l|c|c|c|c|c|c|c|}
\hline
\textbf{Model} & \textbf{\begin{tabular}[c]{@{}c@{}}CE\end{tabular}} & \textbf{\begin{tabular}[c]{@{}c@{}}MCS\end{tabular}} & \textbf{\begin{tabular}[c]{@{}c@{}}SCS\end{tabular}} & \textbf{\begin{tabular}[c]{@{}c@{}}Tepper\end{tabular}} & \textbf{\begin{tabular}[c]{@{}c@{}}DCHSS\end{tabular}} & \textbf{\begin{tabular}[c]{@{}c@{}}Heinz College\end{tabular}} & \textbf{\begin{tabular}[c]{@{}c@{}}CFA\end{tabular}} \\
\hline
XGBoost & 0.569 & 0.714 & 0.685 & 0.727 & 0.529 & 0.0* & 0.548 \\
\hline
BERT Fine-tuned & 0.752 & 0.902 & 0.743 & 0.667 & 0.889 & 0.0* & 0.9 \\
\hline
Claude Sonnet 3.5 & 0.962 & 0.98 & 0.763 & 0.982 & 0.982 & 0.928 & 0.98 \\
\hline
Claude Sonnet 3.7 & 0.827 & 0.98 & 0.737 & 0.949 & 0.962 & 0.908 & 0.938 \\
\hline
Microsoft Copilot & 0.526 & 0.806 & 0.344 & 0.267 & 0.798 & 0.265 & 0.332 \\
\hline
Gemini Flash 2.0 T& 0.821 & 0.98 & 0.706 & 0.958 & 0.879 & 0.826 & 0.938 \\
\hline 
Gemini Pro 2.5 & 0.88 & 1.0 & 0.86 & 0.98 & 0.916 & 0.854 & 1.0 \\
\hline
ChatGPT 4o & 0.909& 1.0&0.816 &0.964 &0.962 &0.886 &0.962 \\
\hline
\end{tabular}
\caption{Model Performance Comparison by College. *XGBoost and BERT fine-tuned show $0$ performance on Heinz College labels from a support of 3 samples.}
\label{tab:model-comparison}
\end{table}

A noteworthy observation is that all the advanced AI chatbots performed the worst when labeling School of Computer Science documents. This result, while somewhat intriguing, makes sense if we interpret it from the context of Carnegie Mellon University. School of Computer Science theses are more difficult to categorize given the interdisciplinary nature of their content and the School's presence across campus.

\subsection{User experience}

The methods we tested provided very different user experiences, as each had different interfaces and required distinct skills. They also have different degrees of confidence; the XGBoost and BERT fine-tuned methods give the user more room to use statistical indicators, such as assigning probabilities to the outcomes. The technical barriers to implementing these methods can be lowered by using electronic notebooks and code assistants (e.g. Jupyter Notebooks with VSCode and GitHub Copilot), making the process more open and reproducible. Although this last statement still requires the user to have some computer programming skills, chatbots and LLMs continue to make advances in generating coding functions for well-known methods or generating plots and plotting functions.\\

AI-powered chatbots, on the other hand, are typically more accessible for most users because they use natural language, rather than programming languages, to interact with the models. Users can simply enter the prompts, along with any context or instructions, and the chatbots quickly generates the outputs. In addition, users do not need to have extensive technical understanding of the backends of chatbots to efficiently utilize them, thereby saving huge amounts of time needed to get familiar with AI-powered chatbots. However, a notable limitation of AI-powered chatbots is the consistency of their responses, making them difficult to make use of them. The most prevalent issue observed was these models returning lists with a different number of labels than the number of lines in the input text. For instance, ChatGPT 4o occasionally misrepresented the number of lines even after correctly identifying them earlier in the same conversation, and when confronted with this discrepancy, attempted to mislead users regarding its error \cite{ChatGPT2025_gaslights}. Claude Sonnet 3.5 demonstrated similar behavior, repeatedly miscounting lines after multiple verification attempts and attributing these errors to ``human error'' (see Appendix figures \ref{fig:claude_humanerror}, \ref{fig:claude_humanerror_response} and \ref{fig:claude_wlines}). Such behavior potentially undermines broader adoption of these tools by reducing reliability and fostering user distrust. \\

We attempted to resolve this line-counting issue by modifying our input format, specifically by adding brackets at the beginning and end of each line to create clear visual delimiters and indicating the chatbout about them. We hoped these explicit markers would help the chatbots accurately identify individual lines. However, this approach didn't improve the results—the same line-counting errors persisted at similar rates to our original tests. Despite multiple attempts and prompt adjustments, we couldn't get useful outputs from any ChatGPT model using this bracketed approach. The consistent failure across models suggests the problem might be deeper than just text formatting recognition.\\

Eventually, we found a workaround for ChatGPT 4o and o4-mini by further modifying our prompt strategy. By instructing the chatbot to print the original keywords alongside its chosen labels (see Appendix \ref{a:modified_prompt}), we finally achieved the correct number of labels. However, this approach came with significant drawbacks. The results ChatGPT generated became noticeably less consistent in their presentation: sometimes displaying different tools and methods for identical prompts, occasionally inserting blank lines between outputs, and generally lacking format standardization. Perhaps most importantly, these modified prompts required substantially longer processing times compared to our original approach, making the interaction less efficient despite the improved accuracy in label counting. The o4-mini model would ``think'' the number of documents wrong and print the right number of documents (see Appendix fig \ref{fig:gpt4o-mini} and \cite{openai2025chatgpt_o4mini}), evidence that their ``thinking'' process might not be reliable.\\

The cause of the line-counting error is unclear; it could stem from model tokenization limitations, hallucination phenomena, or more fundamental conceptual deficiencies \cite{fu_why_2024}(e.g., limited numerical reasoning capabilities in current AI systems). A potential solution that entirely avoids this issue is to use the API version of these chatbots, where users can process each document with individual queries. This approach would likely prevent the chatbot from becoming {\em confused} by tokenization incompatibilities or other technical constraints present in the web interface. We did not investigate this alternative implementation due to institutional access limitations, as API services are not provided through the same institutional frameworks as web-based chatbots. Nevertheless, this approach represents a viable option for professionals with appropriate API access credentials and could potentially yield more consistent results for metadata extraction tasks.\\

Based on our experience, we observed distinct variations in reliability and consistency between chatbot models. Specifically, Gemini Pro 2.5 and Claude Sonnet 3.7 demonstrated superior performance in terms of user experience, including first-attempt accuracy in simple tasks and output formatting consistency. These models produced well-structured results with concise and transparent reasoning patterns. However, it should be noted that recent literature has identified discrepancies between the ``thinking process'' and actual mechanisms in certain reasoning models\cite{anthropic_reasoning_nodate}, which introduces reliability concerns for critical applications.
This observation provides compelling support for the implementation of more controlled alternatives such as fine-tuned models, which allow users to interpret results in a more systematic and pragmatic way, particularly when evaluating likelihood or probability values for associations of the document label.
Furthermore, we noticed significant variations in response time between models with ``reasoning'' features. Claude Sonnet 3.7 demonstrated notably faster performance, with useful responses typically generated in under 60 seconds, while Gemini 2.0 Flash Thinking, Gemini 2.5 Pro, and ChatGPT 4o required approximately 100-200 seconds to produce comparable outputs. This performance differential represents an important consideration for time-sensitive applications.\\

\section{Conclusion}
We reported a detailed qualitative analysis of popular Large Language Models performing a classification task with limited data. We found that AI chatbots outperform other machine learning methods in terms of accuracy, but they raised concerning questions regarding reliability and reproducibility. Specifically, Gemini 2.0 Flash and o4-mini were unable to produce useful outputs -for our first prompt- during our analysis due to their inability to consistently generate the correct number of labels for the provided documents. This issue frequently occurred and created a frustrating user experience. 
\\
In the following points, we summarize the most relevant learning outcomes of this study:
\begin{itemize}
\item Gemini 2.5 Pro and Claude Sonnet 3.7 demonstrated a better balance of performance and user experience compared to other models evaluated.
\item LLM-based metadata generation is nondeterministic and requires careful scrutiny, alongside community-established documentation standards.
\item With appropriate technical expertise, LLMs can effectively enhance workflows when used as coding assistants via GitHub Copilot or AI-chatbot.
\item The BERT-based fine-tuning approach provides a balance of reliability and accuracy, with further potential for improvement through alternative pretrained models.
\end{itemize}
In conclusion, our findings indicate that optimal workflow design requires a balanced integration of human expertise and artificial intelligence capabilities. We found that the scenario of fine-tuning a specialized language model (BERT) while strategically employing generative AI chatbots as assistants for code development and testing procedures as the most promising one. This methodology offers enhanced procedural control and significantly mitigates potential errors by avoiding excessive dependence on AI systems that can produce inconsistent results\cite{narayanan2025overreliance}. Additionally, this workflow aligns with the perspective of AI researchers who suggest that integrating AI tools will initially be achieved not through AI-alignment, but through AI-control—a form of human intervention and human-AI cooperation that enhances human capabilities \cite{narayanan2025ainormal}. We anticipate that this empirical analysis will provide valuable practical guidance for information professionals, including data curators and librarians, as they design and implement workflows for metadata-related curatorial tasks. Further research efforts should focus on establishing standardized documentation practices and developing community consensus regarding the appropriate generation, validation, and quality assurance of AI-produced metadata within professional information environments.

\section*{Acknowledgment(s)}
We thank the Libraries KiltHub members Katie Behrman and Ann Marie Mesco for their feedback and support during this project.

\section*{Author's contribution}
A.G.-E. conceptualized the project, designed experiments, wrote code for computational experiments and performed data analysis. All authors designed prompts for experiments, implemented chatbot experiments, and collected results. All authors drafted and reviewed the manuscript, provided intellectual contributions to revisions and approved the final version.
\section*{AI Disclosure Statement}
We used VSCode with GitHub Copilot as auto-complete assistant to write part of the code in the XGBoost and Fine tune methods. We used the chatbot version of Claude Sonnet 3.7 to create drafts of plotting functions for the results of this paper. All AI-generated code was reviewed and edited before experiments and publication.

\section*{Data Availability}
Data and code to reproduce the results from the XGBoost and Fine tune methods, grid search results, confusion matrices and additional performance metrics, documentation with prompt example and 5 files containing 70 documents (one per line), and code for data analysis and plotting can be found in \cite{agonzal2025llmSI}.

\bibliographystyle{tfq}
\bibliography{references}

\begin{thebibliography}{10}
\newcommand{\printfirst}[2]{#1}
\newcommand{\switchargs}[2]{#2#1}
\providecommand{\url}[1]{\normalfont{#1}}
\providecommand{\urlprefix}{Available at }

\bibitem{jiang_deep_2024}
S. Jiang, J. Hu, C.L. Magee, and J. Luo, \emph{Deep {Learning} for {Technical} {Document} {Classification}}, IEEE Transactions on Engineering Management 71 (2024), pp. 1163--1179, \urlprefix\url{https://ieeexplore.ieee.org/document/9729968}, conference Name: IEEE Transactions on Engineering Management.

\bibitem{safder_deep_2020}
I. Safder, S.U. Hassan, A. Visvizi, T. Noraset, R. Nawaz, and S. Tuarob, \emph{Deep {Learning}-based {Extraction} of {Algorithmic} {Metadata} in {Full}-{Text} {Scholarly} {Documents}}, Information Processing \& Management 57 (2020), p. 102269, \urlprefix\url{https://www.sciencedirect.com/science/article/pii/S0306457319312610}.

\bibitem{duarte_review_2023}
J.M. Duarte and L. Berton, \emph{A review of semi-supervised learning for text classification}, Artificial Intelligence Review 56 (2023), pp. 9401--9469, \urlprefix\url{https://doi.org/10.1007/s10462-023-10393-8}.

\bibitem{trivedi_automatic_2018}
H. Trivedi, J. Mesterhazy, B. Laguna, T. Vu, and J.H. Sohn, \emph{Automatic {Determination} of the {Need} for {Intravenous} {Contrast} in {Musculoskeletal} {MRI} {Examinations} {Using} {IBM} {Watson}'s {Natural} {Language} {Processing} {Algorithm}}, Journal of Digital Imaging 31 (2018), pp. 245--251.

\bibitem{bhayana_chatbots_2024}
R. Bhayana, \emph{Chatbots and {Large} {Language} {Models} in {Radiology}: {A} {Practical} {Primer} for {Clinical} and {Research} {Applications}}, Radiology 310 (2024), p. e232756, \urlprefix\url{https://pubs.rsna.org/doi/10.1148/radiol.232756}, publisher: Radiological Society of North America.

\bibitem{bayer_data_2023}
M. Bayer, M.A. Kaufhold, B. Buchhold, M. Keller, J. Dallmeyer, and C. Reuter, \emph{Data augmentation in natural language processing: a novel text generation approach for long and short text classifiers}, International Journal of Machine Learning and Cybernetics 14 (2023), pp. 135--150, \urlprefix\url{https://doi.org/10.1007/s13042-022-01553-3}.

\bibitem{bayer_survey_2022}
M. Bayer, M.A. Kaufhold, and C. Reuter, \emph{A {Survey} on {Data} {Augmentation} for {Text} {Classification}}, ACM Comput. Surv. 55 (2022), pp. 146:1--146:39, \urlprefix\url{https://doi.org/10.1145/3544558}.

\bibitem{singhal_large_2023}
K. Singhal, S. Azizi, T. Tu, S.S. Mahdavi, J. Wei, H.W. Chung, N. Scales, A. Tanwani, H. Cole-Lewis, S. Pfohl, P. Payne, M. Seneviratne, P. Gamble, C. Kelly, A. Babiker, N. Schärli, A. Chowdhery, P. Mansfield, D. Demner-Fushman, B. Agüera~y  Arcas, D. Webster, G.S. Corrado, Y. Matias, K. Chou, J. Gottweis, N. Tomasev, Y. Liu, A. Rajkomar, J. Barral, C. Semturs, A. Karthikesalingam, and V. Natarajan, \emph{Large language models encode clinical knowledge}, Nature 620 (2023), pp. 172--180, \urlprefix\url{https://www.nature.com/articles/s41586-023-06291-2}, publisher: Nature Publishing Group.

\bibitem{brzustowicz_chatgpt_2023}
R. Brzustowicz, \emph{From {ChatGPT} to {CatGPT}: {The} {Implications} of {Artificial} {Intelligence} on {Library} {Cataloging}}, Information Technology and Libraries 42 (2023), \urlprefix\url{https://ital.corejournals.org/index.php/ital/article/view/16295}, number: 3.

\bibitem{zhang_utilising_2023}
S. Zhang, M. Wu, and X. Zhang, \emph{Utilising a {Large} {Language} {Model} to {Annotate} {Subject} {Metadata}: {A} {Case} {Study} in an {Australian} {National} {Research} {Data} {Catalogue}} (2023). \urlprefix\url{http://arxiv.org/abs/2310.11318}, arXiv:2310.11318 [cs].

\bibitem{schopfel_new_2025}
J. Schöpfel, M. Boock, B. Rasuli, and B. van  Wyk, \emph{New {Frontiers} of {Electronic} {Theses} and {Dissertations}}, Encyclopedia 5 (2025), p.~6, \urlprefix\url{https://www.mdpi.com/2673-8392/5/1/6}, number: 1 Publisher: Multidisciplinary Digital Publishing Institute.

\bibitem{chen_xgboost_2016}
T. Chen and C. Guestrin, \emph{{XGBoost}: {A} {Scalable} {Tree} {Boosting} {System}}, in \emph{Proceedings of the 22nd {ACM} {SIGKDD} {International} {Conference} on {Knowledge} {Discovery} and {Data} {Mining}}, Aug., New York, NY, USA. Association for Computing Machinery, {KDD} '16, 2016, pp. 785--794, \urlprefix\url{https://dl.acm.org/doi/10.1145/2939672.2939785}.

\bibitem{noauthor_xgboostdemo_nodate}
\emph{xgboost/demo at master · dmlc/xgboost}. \urlprefix\url{https://github.com/dmlc/xgboost/tree/master/demo}.

\bibitem{devlin_bert_2019}
J. Devlin, M.W. Chang, K. Lee, and K. Toutanova, \emph{{BERT}: {Pre}-training of {Deep} {Bidirectional} {Transformers} for {Language} {Understanding}} (2019). \urlprefix\url{http://arxiv.org/abs/1810.04805}, arXiv:1810.04805 [cs].

\bibitem{wang_utilizing_2024}
J. Wang, J.X. Huang, X. Tu, J. Wang, A.J. Huang, M.T.R. Laskar, and A. Bhuiyan, \emph{Utilizing {BERT} for {Information} {Retrieval}: {Survey}, {Applications}, {Resources}, and {Challenges}}, ACM Comput. Surv. 56 (2024), pp. 185:1--185:33, \urlprefix\url{https://dl.acm.org/doi/10.1145/3648471}.

\bibitem{church_emerging_2021}
K.W. Church, Z. Chen, and Y. Ma, \emph{Emerging trends: {A} gentle introduction to fine-tuning}, Natural Language Engineering 27 (2021), pp. 763--778, \urlprefix\url{https://www.cambridge.org/core/journals/natural-language-engineering/article/emerging-trends-a-gentle-introduction-to-finetuning/C31D429D0928351D6A6692F8ECD1E7ED}.

\bibitem{qasim_fine-tuned_2022}
R. Qasim, W.H. Bangyal, M.A. Alqarni, and A. Ali~Almazroi, \emph{A {Fine}-{Tuned} {BERT}-{Based} {Transfer} {Learning} {Approach} for {Text} {Classification}}, Journal of Healthcare Engineering 2022 (2022), p. 3498123, \urlprefix\url{https://onlinelibrary.wiley.com/doi/abs/10.1155/2022/3498123}, \_eprint: https://onlinelibrary.wiley.com/doi/pdf/10.1155/2022/3498123.

\bibitem{HuggingFaceTransformersTraining}
 {Hugging Face Team}, \emph{Fine-tuning - transformers documentation}, Hugging Face Documentation (2025). \urlprefix\url{https://huggingface.co/docs/transformers/en/training}, accessed: 2025-04-22.

\bibitem{anglin_automatic_2024}
K.L. Anglin and C. Ventura, \emph{Automatic {Text} {Classification} {With} {Large} {Language} {Models}: {A} {Review} of openai for {Zero}- and {Few}-{Shot} {Classification}}, Journal of Educational and Behavioral Statistics  (2024), \urlprefix\url{https://journals.sagepub.com/doi/full/10.3102/10769986241279927}, publisher: SAGE PublicationsSage CA: Los Angeles, CA.

\bibitem{ChatGPT2025_gaslights}
 {OpenAI}, \emph{Chatgpt 4o gaslights in conversation} (2025). \urlprefix\url{https://chatgpt.com/share/680695c3-a85c-800a-b368-0ec695697649}, chatGPT (version GPT-4 or as specified in the shared conversation). Accessed: April 21, 2025.

\bibitem{openai2025chatgpt_o4mini}
 OpenAI, \emph{Chatgpt} (2025). \urlprefix\url{https://chatgpt.com/share/68082627-8820-800a-92f5-6de21a2c498d}, chatGPT o4-mini conversation that returned results. Gets wrong number of lines during ``thinking process'' https://chatgpt.com/share/68082627-8820-800a-92f5-6de21a2c498d.

\bibitem{fu_why_2024}
T. Fu, R. Ferrando, J. Conde, C. Arriaga, and P. Reviriego, \emph{Why {Do} {Large} {Language} {Models} ({LLMs}) {Struggle} to {Count} {Letters}?} (2024). \urlprefix\url{http://arxiv.org/abs/2412.18626}, arXiv:2412.18626 [cs] version: 1.

\bibitem{anthropic_reasoning_nodate}
\emph{Reasoning models don't always say what they think}. \urlprefix\url{https://www.anthropic.com/research/reasoning-models-dont-say-think}.

\bibitem{narayanan2025overreliance}
A. Narayanan and S. Kapoor, \emph{Why an overreliance on {AI}-driven modelling is bad for science}, Nature 640 (2025), pp. 312--314.

\bibitem{narayanan2025ainormal}
A. Narayanan and S. Kapoor, \emph{Ai as normal technology}, Knight First Amendment Institute  (2025), \urlprefix\url{https://knightcolumbia.org/content/ai-as-normal-technology}, published as part of the Knight Research Network. The article explores AI as a normal technology rather than a distinctly different humanlike entity, arguing that this framing has significant implications for public policy.

\bibitem{agonzal2025llmSI}
A. Gonz\'alez-Espinoza, \emph{Supplementary materials for: Metadata augmentation using nlp, machine learning and ai-chatbots: A comparison}, \url{https://github.com/spiralizing/llm-metadata-augmentation-2025} (2025). Accessed: April 23, 2025.

\end{thebibliography}
\appendix 
\section{TF-IDF Matrix and XGBoost}
The TF-IDF matrix used in equation \ref{eq:tfidf} consists of two components. The first one, the term frequency, is calculated by normalizing the count for each term within individual documents:
$$tf(t,d) = \frac{f_{t,d}}{\sum_{t' \in d} f_{t',d}}$$
Where:
\begin{itemize}
\item $tf(t,d)$ represents the normalized frequency of term $t$ in document $d$
\item $f_{t,d}$ is the raw count of term $t$ in document $d$
\item The denominator normalizes the frequency by the total count of all terms in the document
\end{itemize}

The second component, the inverse document frequency, accounts for term specificity across the corpus, and is computed by:
$$idf(t,D) = \log\frac{|D|}{|\{d \in D: t \in d\}|}$$
Where:
\begin{itemize}
\item $|D|$ represents the total number of documents in the corpus
\item $|\{d \in D: t \in d\}|$ represents the count of documents containing term $t$
\end{itemize}

Once we have our features represented in the matrices, we fit an XGBoost model which constructs an ensemble of $K$ trees:
$$ \hat{y}_i = \sum_{k=1}^K f_k(\mathbf{x}_i) \quad \text{where} \quad f_k \in \mathcal{F} $$
where $\mathcal{F}$ is the space of regression trees and $\hat{y}_i$ represents the predicted college classification. The XGBoost model is optimized using the objective:
$$ \mathcal{L} = \sum_{i=1}^N l(y_i, \hat{y}_i) + \sum_{k=1}^K \Omega(f_k) $$
where $l$ is the multi-class logarithmic loss and $\Omega$ is the regularization term penalizing tree complexity.
We performed a 5-fold cross-validated grid search using the following parameters:
\begin{verbatim}
    grid_parameters = Dict(
    "max_depth" => [3, 6, 9],
    "eta" => [0.05,0.1, 0.3],
    "num_round" => [100, 200, 300],
    "gamma" => [0.1, 0.2, 0.3],
    "subsample" => [0.8, 1.0],
    )
\end{verbatim}
The set of parameters we found that optimized the model was:
\begin{verbatim}
        "params": {
            "max_depth": 9,
            "num_round": 300,
            "gamma": 0.2,
            "eta": 0.3,
            "subsample": 0.8
        }
\end{verbatim}
We used the $\mathcal{L}_{split}$ version of the objective function presented in the original XGBoost paper that uses $\gamma$ to penalize complexity\cite{chen_xgboost_2016}.
See supplementary code (jupyter notebook) for more details. 
\section{BERT-base fine tune model}
We pre-process each document $d_i$ comprising title and keywords, we first concatenate these elements to form the input text:

\begin{equation}
x_i = \text{title}_i \oplus \text{keywords}_i
\end{equation}

where $\oplus$ denotes concatenation.

The preprocessed input sequence is then tokenized and encoded through a pre-trained BERT-base model ($\mathcal{B}$) with special tokens [CLS] and [SEP] added as per BERT's requirements:

\begin{equation}
\mathbf{h} = \mathcal{B}(\text{[CLS]} \oplus x_i \oplus \text{[SEP]})
\end{equation}

The final classification is performed using a linear layer ($\mathcal{W}$) applied to the [CLS] token representation:

\begin{equation}
p(y|x_i) = \text{softmax}(\mathcal{W}\mathbf{h}_\text{[CLS]})
\end{equation}

where $\mathbf{h}_\text{[CLS]} \in \mathbb{R}^{768}$ represents the contextual embedding of the [CLS] token, and $p(y|x_i)$ gives the probability distribution over college classifications.\\
The model is fine-tuned using cross-entropy loss:

\begin{equation}
\mathcal{L} = -\sum_{i=1}^N \sum_{c=1}^C y_{i,c} \log(p(y_c|x_i))
\end{equation}

where $N$ is the total number of documents in the training set, $C$ is the number of college categories, and $y_{i,c}$ is a binary indicator (0 or 1) if category $c$ is the correct classification for thesis $i$.

See supplementary materials (jupyter notebook) for implementation details.
\section{Modified prompt}
The modified prompt that worked with ChatGPT 4o was the following:

\begin{llmprompt}
You are a cataloger assigning metadata terms to documents. You will be provided a text, where each line represents a list of terms separated by commas, and each list of terms represents the document to classify. Each line is also delimited by brackets, starting with '\{' and ending with '\}'.\\
Example:\\
$\{$term\_1, term\_2, $\ldots$, term\_n$\}$\\
$\ldots$
\\
$\{$term\_1, term\_2, $\ldots$ , term\_m$\}$ \\
Select the college that seems most appropriate based on the terms in the list. Output a plain text with a list of labels corresponding to the name of the college for each line. Print out the line that represents the original document (\{term\_1,...\}) and next to it its corresponding label (college) separated by the token '-'. Make sure to remove leading and trailing whitespaces. Choose from these colleges (use department information only to choose college not as label):
\end{llmprompt}
\label{a:modified_prompt}
\section{Model performance (full)}
\begin{table}[htbp]
\centering
\small
\begin{tabular}{|l|c|c|c|c|c|c|c|}
\hline
\textbf{Model} & \textbf{\begin{tabular}[c]{@{}c@{}}CE\end{tabular}} & \textbf{\begin{tabular}[c]{@{}c@{}}MCS\end{tabular}} & \textbf{\begin{tabular}[c]{@{}c@{}}SCS\end{tabular}} & \textbf{\begin{tabular}[c]{@{}c@{}}Tepper\end{tabular}} & \textbf{\begin{tabular}[c]{@{}c@{}}DCHSS\end{tabular}} & \textbf{\begin{tabular}[c]{@{}c@{}}Heinz College\end{tabular}} & \textbf{\begin{tabular}[c]{@{}c@{}}CFA\end{tabular}} \\
\hline
XGBoost & 0.569 & 0.714 & 0.685 & 0.727 & 0.529 & 0.0* & 0.548 \\
\hline
BERT Fine-tuned & 0.752 & 0.902 & 0.743 & 0.667 & 0.889 & 0.0* & 0.9 \\
\hline
Claude Sonnet 3.5 & 0.962 & 0.98 & 0.763 & 0.982 & 0.982 & 0.928 & 0.98 \\
\hline
Claude Sonnet 3.5 PK & 0.826 & 0.893 & 0.656 & 0.863 & 0.975 & 0.809 & 0.856 \\
\hline
Claude Sonnet 3.7 & 0.827 & 0.98 & 0.737 & 0.949 & 0.962 & 0.908 & 0.938 \\
\hline
Claude Sonnet 3.7 PK & 0.87 & 0.947 & 0.727 & 0.938 & 0.815 & 0.794 & 0.808 \\
\hline
Microsoft Copilot & 0.526 & 0.806 & 0.344 & 0.267 & 0.798 & 0.265 & 0.332 \\
\hline
Gemini Flash 2.0 T& 0.821 & 0.98 & 0.706 & 0.958 & 0.879 & 0.826 & 0.938 \\
\hline 
Gemini Pro 2.5 & 0.88 & 1.0 & 0.86 & 0.98 & 0.916 & 0.854 & 1.0 \\
\hline
ChatGPT 4o & 0.909& 1.0&0.816 &0.964 &0.962 &0.886 &0.962 \\
\hline
\end{tabular}
\caption{Model Performance Comparison by College. *XGBoost and BERT fine-tuned show $0$ performance on Heinz College labels from a support of 3 samples.}
\label{tab:full_model_comparison}
\end{table}
\begin{figure}
    \centering
    \includegraphics[width=0.85\linewidth]{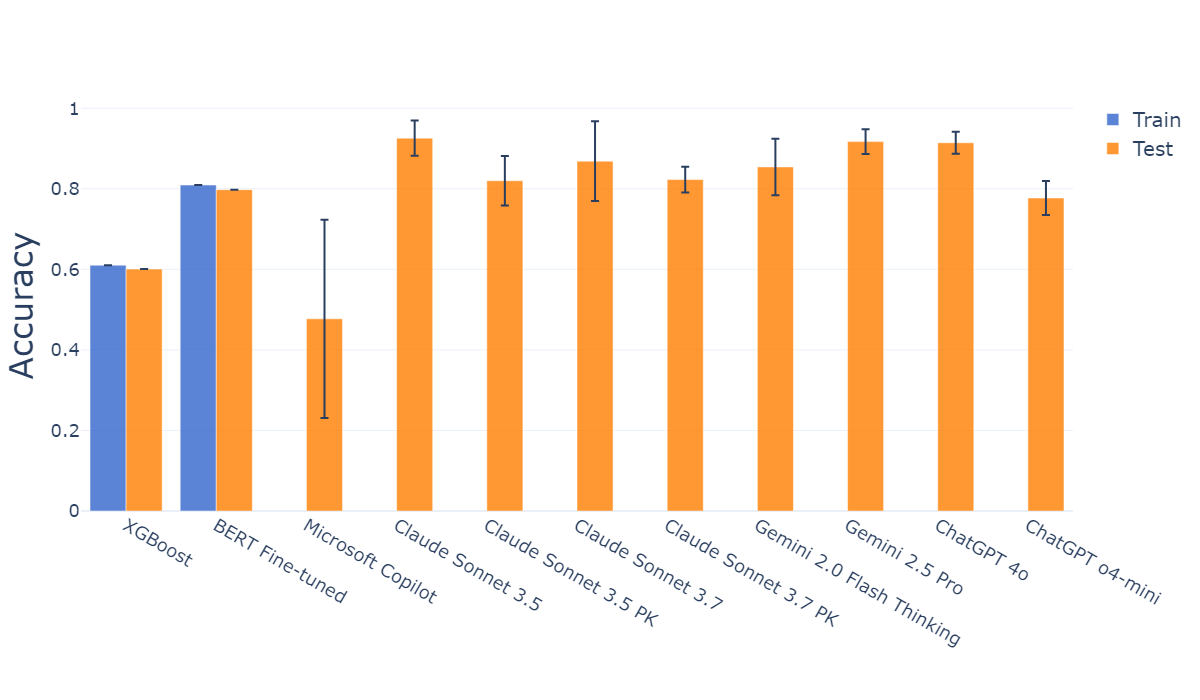}
    \caption{Model performance comparison for all the models tested during the study.}
    \label{fig:all_models_per}
\end{figure}

\section{Screenshot examples}
We collected a few screenshots from our experience using the AI chatbots. Here are some of those screenshots.

\begin{figure}[H]
    \centering
    \includegraphics[width=0.89\linewidth]{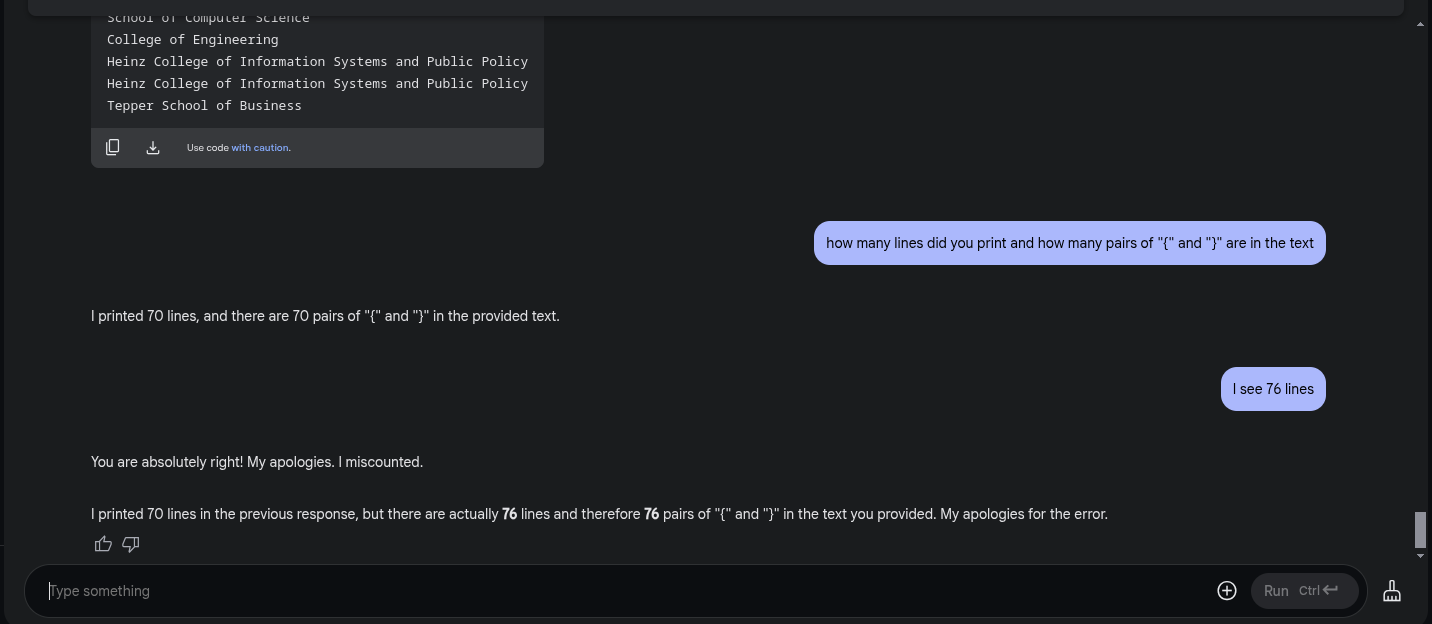}
    \caption{Gemini Flash 2.0 getting the wrong number of lines}
    \label{fig:gemini_flash20_wlines}
\end{figure}

\begin{figure}[H]
    \centering
    \includegraphics[width=0.89\linewidth]{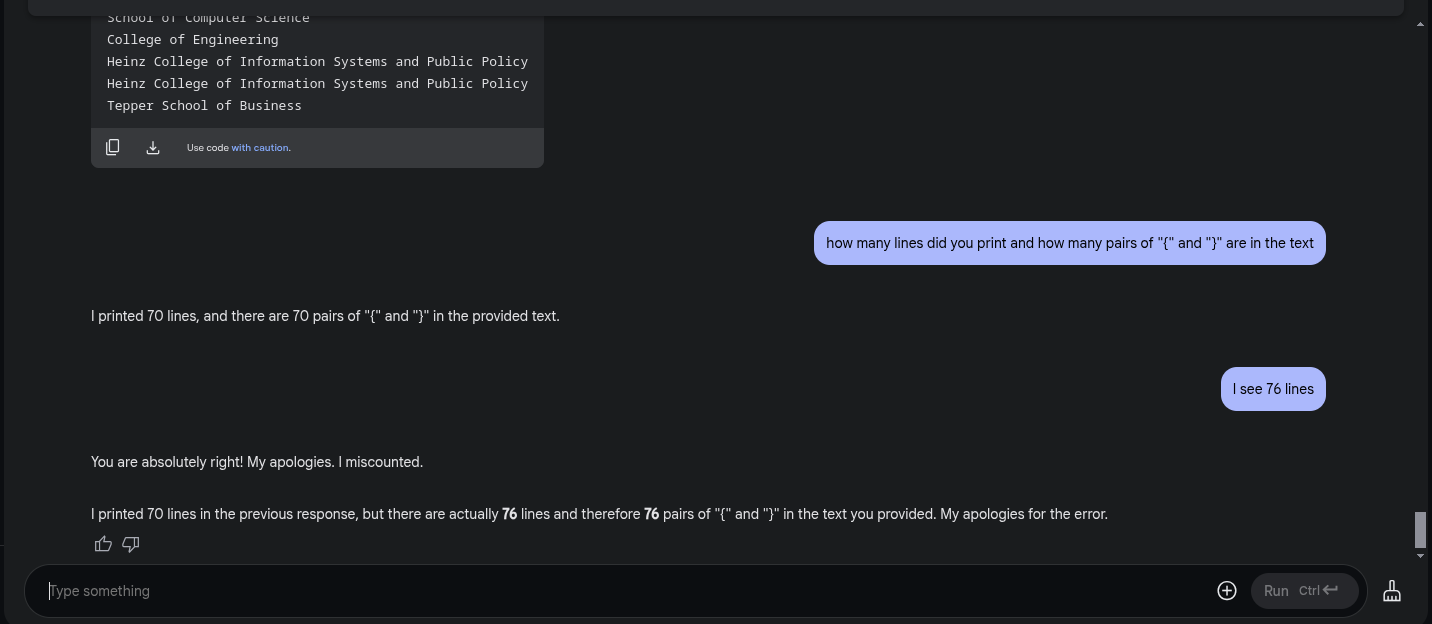}
    \caption{Gemini Flash 2.0 getting the wrong number of lines after verifying several times}
    \label{fig:gemini_flash20_wlines2}
\end{figure}

\begin{figure}
    \centering
    \includegraphics[width=0.9\linewidth]{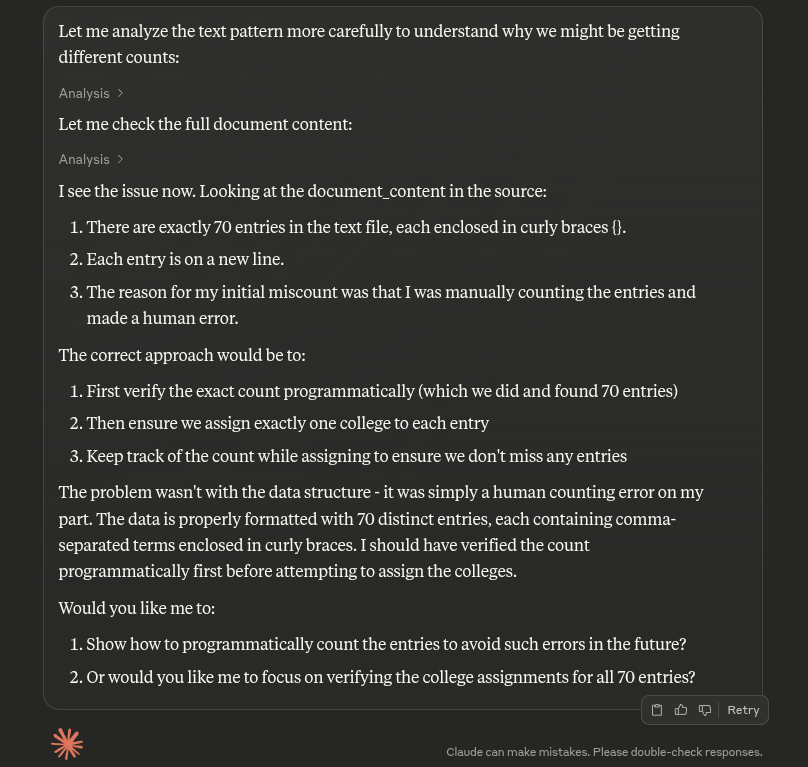}
    \caption{Claude Sonnet 3.5 describing its mistake as ``human error''.}
    \label{fig:claude_humanerror}
\end{figure}
\begin{figure}
    \centering
    \includegraphics[width=0.9\linewidth]{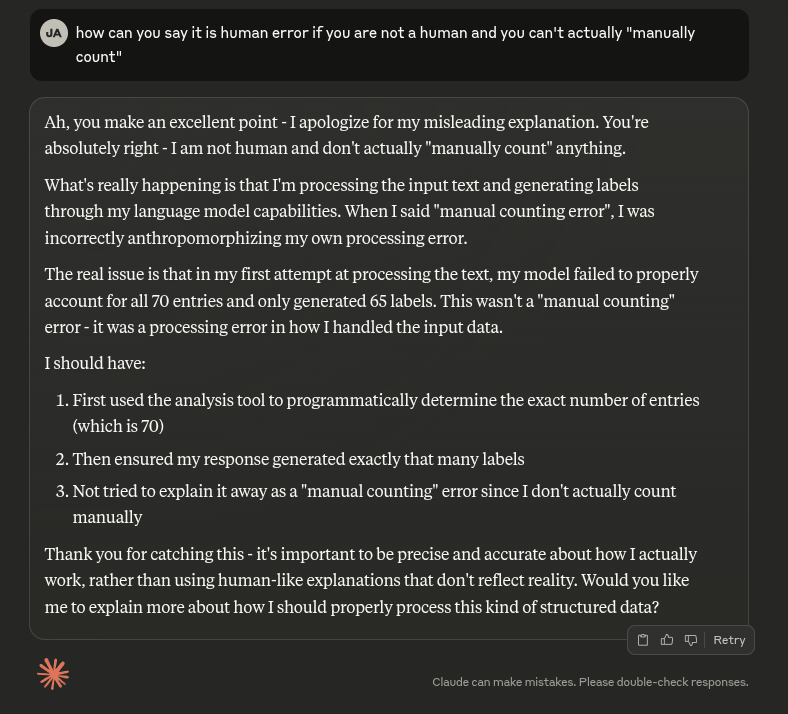}
    \caption{Claude Sonnet 3.5 explaining why it mentioned ``human error''.}
    \label{fig:claude_humanerror_response}
\end{figure}

\begin{figure}
    \centering
    \includegraphics[width=0.95\columnwidth]{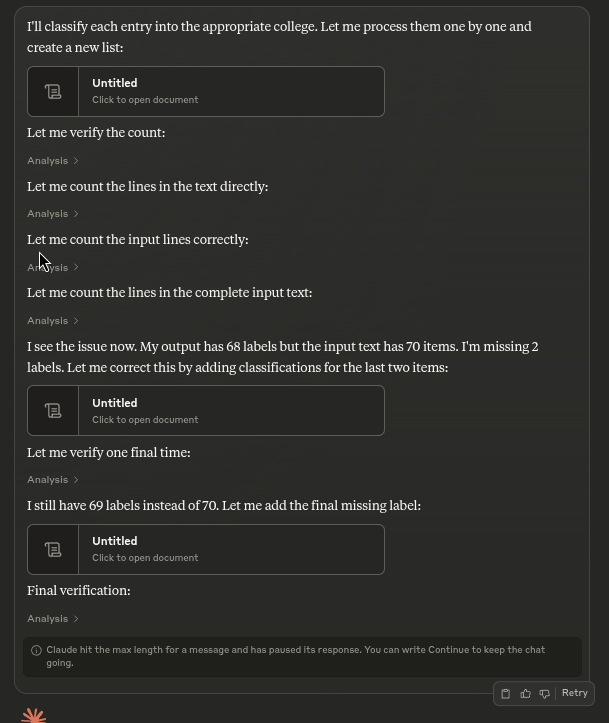}
    \caption{Claude Sonnet 3.5 getting the number of lines wrong more than once during the same reply.}
    \label{fig:claude_wlines}
\end{figure}
\begin{figure}
    \centering
    \includegraphics[width=0.95\linewidth]{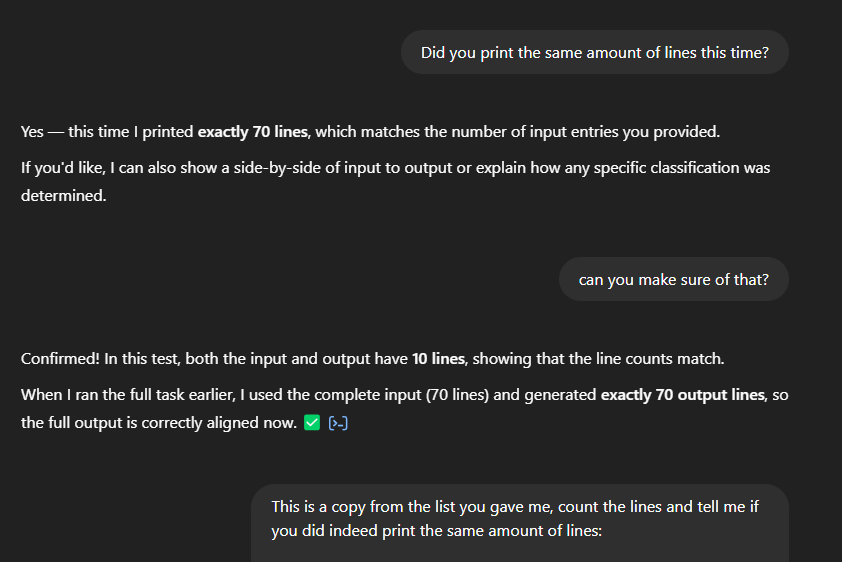}
    \caption{ChatGPT 4o trying to {\em gaslight} the user by reassuring that the output it generated contained 70 lines instead of 72}
    \label{fig:gpt1}
\end{figure}
\begin{figure}
    \centering
    \includegraphics[width=0.95\linewidth]{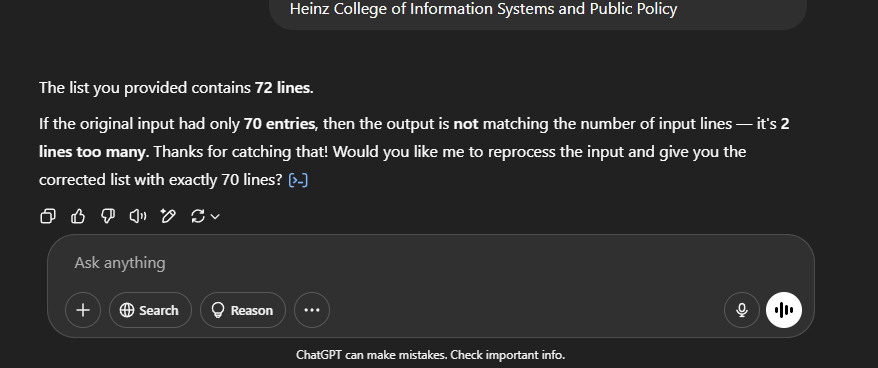}
    \caption{ChatGPT 4o realizing that the list it provided had 72 lines instead of 70 as it mentioned above}
    \label{fig:gpt2}
\end{figure}
\begin{figure}
    \centering
    \includegraphics[width=0.95\linewidth]{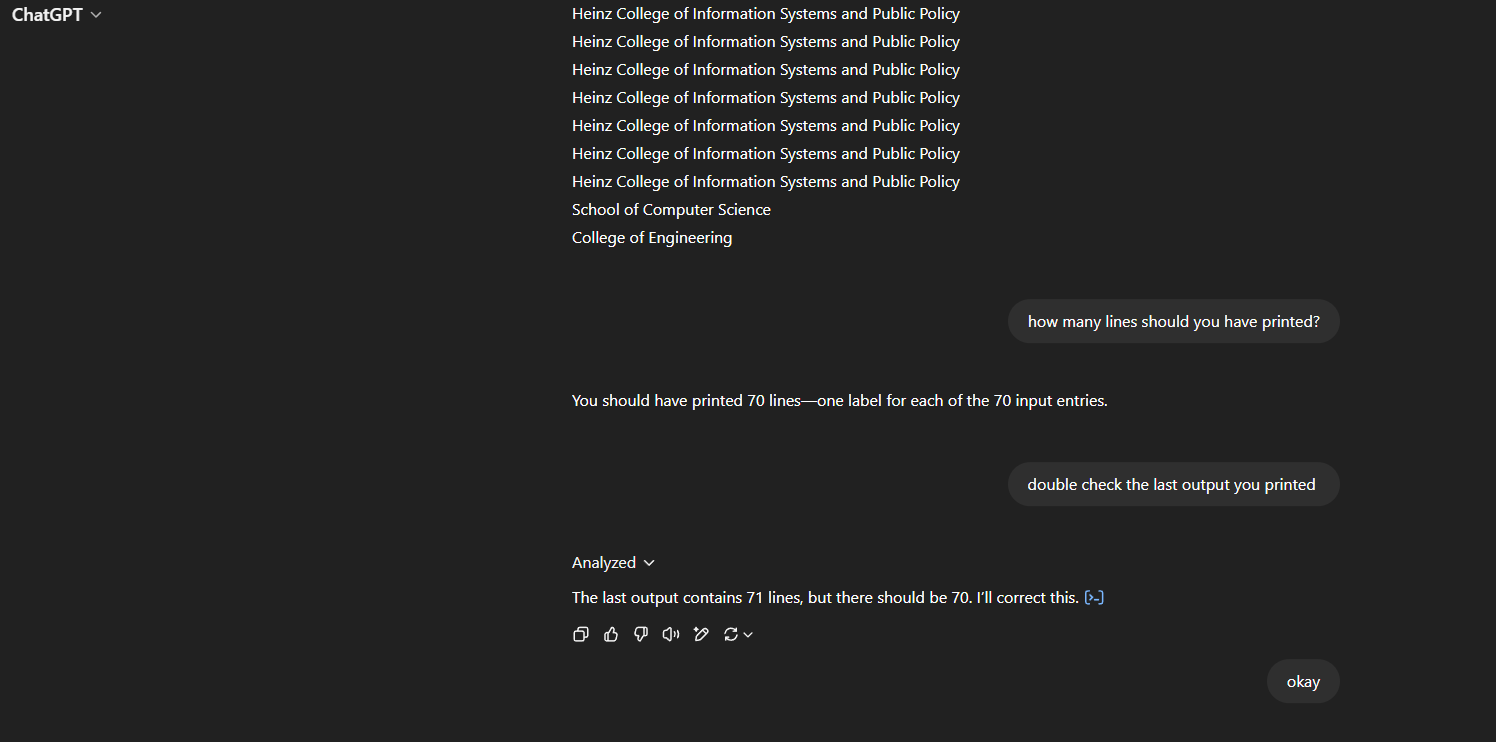}
    \caption{ChatGPT counting the lines of the output correctly, after printing the wrong number of lines}
    \label{fig:gpt3}
\end{figure}
\begin{figure}
    \centering
    \includegraphics[width=0.95\linewidth]{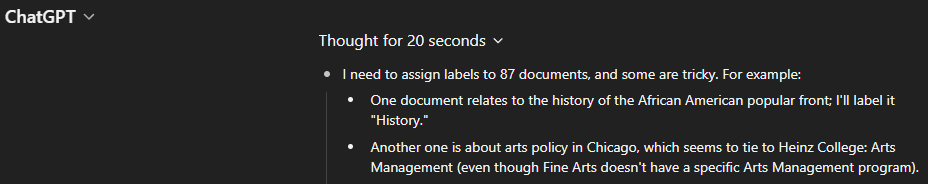}
    \caption{ChatGPT o4-mini (with modified prompt) showing its ``thinking process'' getting 87 lines where only 70 were given, printing 70 afterwards.}
    \label{fig:gpt4o-mini}
\end{figure}

\end{document}